\documentclass[journal=jacsat,manuscript=article]{achemso}

\usepackage{lineno}
\usepackage{tabularx}
\usepackage{amsmath}
\usepackage[caption=false]{subfig}

\newcommand\clearrow{\global\let\rowmac\relax}
\clearrow

\title{Sub-2 Kelvin characterization of nitrogen-vacancy centers in silicon carbide nanopillars}

\author{Victoria A. Norman}
\email{vanorman@ucdavis.edu}
\affiliation[ECE]{Department of Electrical and Computer Engineering, University of California, Davis, Davis, CA 95620, USA}
\alsoaffiliation[Phys]{Department of Physics and Astronomy, University of California Davis, Davis, CA 95620, USA}
\author{Sridhar Majety}
\author{Alex H. Rubin}
\alsoaffiliation[Phys]{Department of Physics and Astronomy, University of California Davis, Davis, CA 95620, USA}

\author{Pranta Saha}

\author{Nathan R. Gonzalez}
\alsoaffiliation[Phys]{Department of Physics and Astronomy, University of California Davis, Davis, CA 95620, USA}

\author{Jeanette Simo}

\author{Bradi Palomarez}

\author{Liang Li}
\alsoaffiliation[Phys]{Department of Physics and Astronomy, University of California Davis, Davis, CA 95620, USA}
\affiliation[ECE]{Department of Electrical and Computer Engineering, University of California, Davis, Davis, CA 95620, USA}

\author{Pietra B. Curro}
\affiliation[SCU]{Department of Mechanical Engineering, Santa Clara University, 500 El Camino Real, Santa Clara, CA 95053, USA}

\author{Scott Dhuey}
\author{Selven Virasawmy}
\affiliation[MF]{The Molecular Foundry, Lawrence Berkeley National Laboratory, 1 Cyclotron Rd, Berkeley, CA 94720, USA}

\author{Marina Radulaski}
\affiliation[ECE]{Department of Electrical and Computer Engineering, University of California, Davis, Davis, CA 95620, USA}
\email{mradulaski@ucdavis.edu}

\begin{document}

\begin{abstract} 

The development of efficient quantum communication technologies depends on the innovation in multiple layers of its implementation, a challenge we address from the fundamental properties of the physical system at the nano-scale to the instrumentation level at the macro-scale. We select a promising near infrared quantum emitter, the nitrogen-vacancy (NV) center in 4H-SiC, and integrate it, at an ensemble level, with nanopillar structures that enhance photon collection efficiency into an objective lens. Moreover, changes in collection efficiency in pillars compared to bulk can serve as indicators of color center orientation in the lattice. To characterize NV center properties at the unprecedented sub-2 Kelvin temperatures, we incorporate compatible superconducting nanowire single photon detectors inside the chamber of an optical cryostat and create the ICECAP, the Integrated Cryogenic system for Emission, Collection And Photon-detection. ICECAP measurements show no significant linewidth broadening of NV ensemble emission and up to 14-fold enhancement in collected emission. With additional filtering, we measure emitter lifetimes of NV centers in a basal ($hk$) and an axial ($kk$) orientation unveiling their cryogenic values of 2.2 ns and 2.8 ns.

\end{abstract}

\section{Introduction}

Quantum emitters in semiconductor substrates is a rapidly maturing field for quantum information technologies. However, many of the ambitious plans for nanophotonic-based quantum computers require significant work on the integration of on-chip single-photon sources and detectors  \cite{moody_2022_2022}. There is an abundance of implementations of quantum emitters that can be integrated with nanophotonic devices: semiconductor quantum dots, color centers in diamond, silicon, and silicon carbide, and more recently in 2D materials like hBN or CrCl$_3$ as well as rare earth ions implanted into semiconductors \cite{becher20232023, kianinia2022quantum, castelletto2022silicon, uppu2021quantum, elshaari2020hybrid}.
In particular, silicon carbide has a number of useful optical properties such as the large electronic bandgap of $3.26$ eV that allows for a wide variety of color centers and the availability of high quality, affordable, large-scale wafers and as such has attracted attention as a quantum information platform \cite{Khazen2023, Weber2010, PhysRevB.94.121202, majety2022quantum}.
Specifically, the NV center in 4H-SiC has been characterized at room temperature as a single-photon source with a lifetime of 2.1 - 2.8 ns \cite{JFWangCoherentcontrol, wang2020coherent}. Additionally, coherent control of NV centers has been demonstrated on both ensembles and single emitters at room temperature and 10 K \cite{JFWangCoherentcontrol, ZMuCoherentManip}. Further, work is being done to enable the transfer of electron-spin polarization to nearby nuclear spins \cite{Murzakhanov2023} which could enable quantum memory technology. Single digit temperature characterization and integration into nanophotonic devices are two areas that had previously remained unexplored in the NV center parameter space until this work.

In parallel with demonstrations of the 4H-SiC NV center's  potential for quantum communication, two significant developments in commercially available lab equipment occurred: closed-cycle, optical cryostats that reliably achieve temperatures below 2 K from companies like Montana Instruments and ICEoxford, and superconducting nanowire single-photon detectors (SNSPDs) that have high sensitivity ($\geq 80 \%$), low dark counts ($< 100$ Hz), and low recovery time ($\sim 50$ ns) with operating temperatures above 2 K from companies like Quantum Opus and Single Quantum. The potential combination of these two technologies has led to new opportunities for quantum optics groups that work in the single-photon regime. There have been experiments using \textit{in situ} custom SNSPDs\cite{PhysRevApplied.19.014037} but these systems do not have optical access, cannot use confocal techniques, require special sample preparation methods, and do not appear to be compatible with photonic integration. We introduce the ICECAP, an Integrated Cryogenic system for Emission, Collection And Photon-detection, as the first such deployable incorporated system. We then use it to demonstrate low single-digit temperature spectral properties of NV centers integrated into 4H-SiC nanopillars for the first time. This work represents a significant step toward the NV center in 4H-SiC as a scalable, telecom-compatible, nanophotonic platform.

\section{Modeling of color center emission collection}
\label{section:fab}
Color centers in bulk emit light in all directions which results in significant losses when photons are collected via an objective lens. For the purposes of quantum information processing, enhanced collection, achievable by, for example, integration with, nanopillars, would impact the color center's rate as a single-photon source, entanglement distribution success rate, as well as the sensitivity of the spin-qubit readout.

\begin{figure}[htbp]
    \centering
    \includegraphics[width = \linewidth]{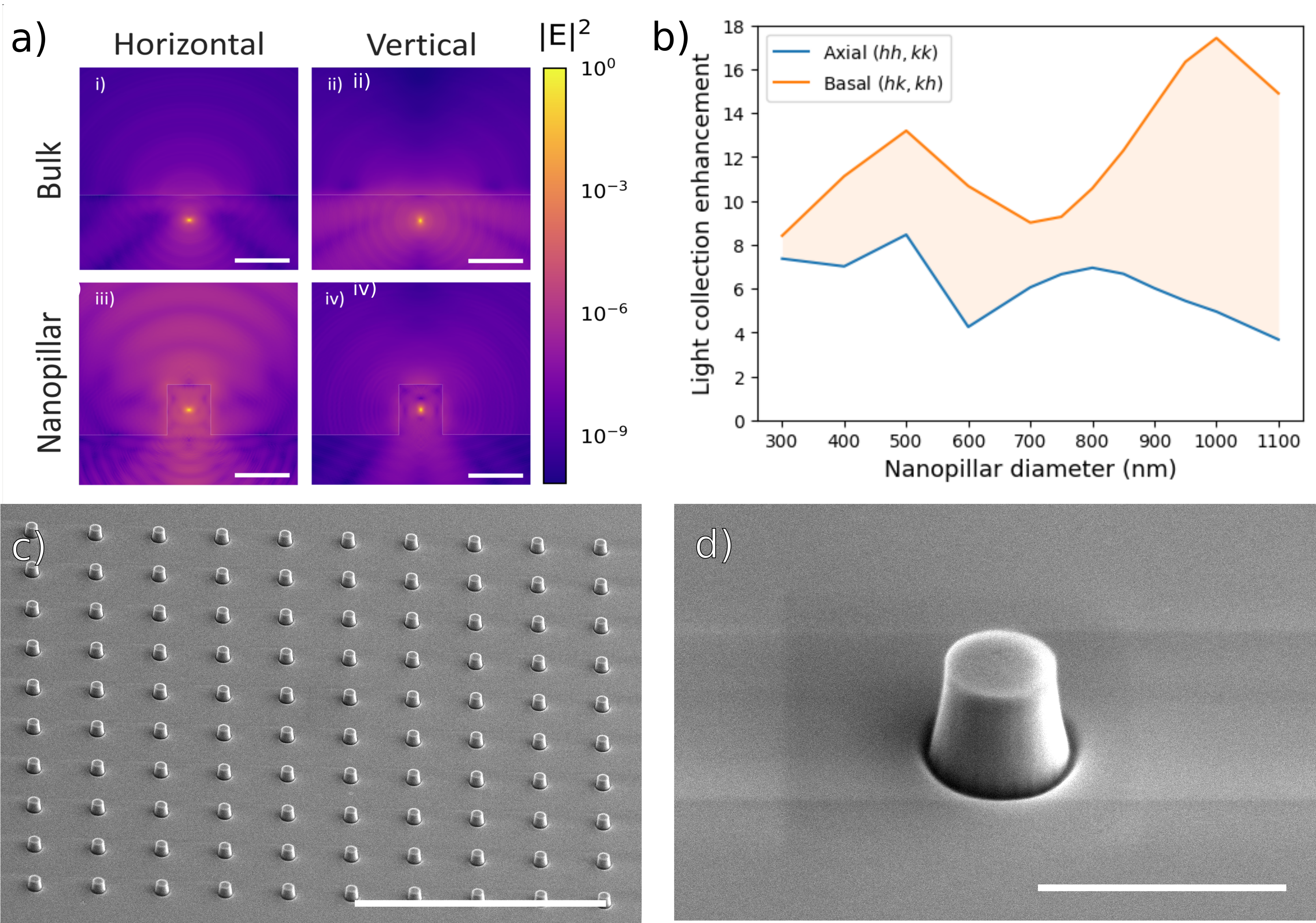}
    \caption{a) Simulated electric field intensity profiles of a (i,iii) horizontally-  and (ii,iv) vertically-oriented electric dipole emission from (i, ii) bulk and (iii, iv) the center of a 800 nm diameter nanopillar. All four plots are normalized to the same value, b) Simulated values of light collection efficiency enhancement as a function of the nanopillar diameter, for different orientations of the NV dipole. Collection for axial NV centers are based on emission from a dipole oriented perpendicularly to the c-axis;
    basal center collection represents the enhancement range over all possible dipole orientations, from 19$^\circ$ to 90$^\circ$ relative to the c-axis, as discussed in the main text, c) SEM image of an array of 850 nm diameter SiC nanopillars (scale bar represents 20 $\mu$m), d) SEM image of a single 850 nm diameter SiC nanopillar (scale bar represents 2 $\mu$m). }
    \label{fig:Fig2}
\end{figure}

We use the finite-difference time-domain (FDTD) package in Ansys Lumerical to simulate the outcoupling properties of SiC nanopillars at the approximate emission wavelength of the NV color center in 4H-SiC. Nanopillars with a height of 1 $\mu$m and diameters varying from 300-1100 nm are simulated, with a dipole emitting at 1300 nm and positioned at the lateral and longitudinal center of the nanopillar, and compared to the case when the dipole is at placed 500 nm under the surface of bulk. Collection efficiency for bulk is simulated to be 3.7\% for horizontally oriented dipole and 0.63\% for vertically-oriented dipole. We calculate the light collection efficiency of the color center emission for an objective with numerical aperture of 0.85, positioned above the nanopillar or bulk models. We observe that the nanopillar guides more of the dipole emission towards the objective, especially for the vertically oriented dipole, as shown in Figure~\ref{fig:Fig2}a. To connect these simulation results to NV center emission, we first need to consider the nontrivial structure of this emitter's dipole orientation. The NV center possesses two degenerate transition dipole moments, which lie in the plane perpendicular to the N-V axis (the line connecting the nitrogen and the vacant lattice site) \cite{PhysRevLett.101.226403, PhysRevB.109.235203}.
For axial centers, whose N-V axis is parallel to the crystal c-axis, all emission is from a dipole oriented perpendicular to the c-axis (horizontal dipole orientation). Modeled collection efficiency for varying pillar diameters in Fig~\ref{fig:Fig2}b (blue line) shows up to 8-fold enhancement in collected light for axial NV centers, compared to bulk. 
For basal centers, whose N-V axis lies at 71$^\circ$ relative to the c-axis \cite{talwar2023structure}, the dipole orientation can be anywhere from perpendicular to the c-axis to 19$^\circ$ off the c-axis, represented by the orange range of values in Fig~\ref{fig:Fig2}b with up to 18-fold enhancement.
It is not currently known if the basal defects have a preferred discrete angular dipole orientation.

The collection efficiency $c(\beta, d)$ of a dipole oriented at angle $\beta$ relative to the c-axis is calculated as

  $  c(\beta, d) = c(0^\circ, d) \cos^2(\beta) + c(90^\circ, d) \sin^2(\beta)$,

where $c(0^\circ, d)$ and $c(90^\circ, d)$ are the collection efficiencies for dipoles parallel and perpendicular to the c-axis, respectively, as determined from the FDTD simulations.

\section{Fabrication of SiC quantum photonic devices}

To fabricate pillar samples, a HPSI 4H-SiC wafer is commercially implanted (CuttingEdge Ions, LLC) with $^{14}$N$^{+}$ ions at an energy of 375 keV and a dose of 1 $\times$ 10$^{14}$ cm$^{-2}$ \cite{wang2020coherent}. As simulated by Stopping and Range of Ions in Matter (SRIM), this energy is expected to generate a peak nitrogen concentration at a depth of 500 nm. The implants are subsequently activated by annealing the samples in a 1-inch Lindberg Blue tube furnace at 1050 $^{\circ}$C in nitrogen atmosphere for 60 minutes \cite{wang2020coherent}. Arrays of circular holes with diameters ranging from 300-1100 nm are then patterned onto a 350 nm PMMA layer using e-beam lithography. A 5 nm titanium adhesion layer followed by a 50 nm nickel hard mask layer are then e-beam evaporated and lifted off to transfer the circular patterns to this metal layer. Nanopillars are etched into SiC through inductively coupled plasma reactive ion etching (ICP-RIE) involving SF$_6$ and O$_2$ chemistry \cite{radulaski2017scalable}. The SF$_6$ and O$_2$ flow rates are maintained at a ratio of 4:1 to achieve an etch rate of $\sim$300 nm/min and a pillar height of 1 $\mu$m. The metal layer is then removed by etching the samples in Transene's nickel and titanium etchants. Scanning electron microscope (SEM) images of the fabricated nanopillars are shown in Figure~\ref{fig:Fig2}c-d.

\section{An optical cryostat with incorporated single-photon detectors}
\label{sec:instrument}
\begin{figure}[htbp]
    \centering
    \includegraphics[width = \linewidth]{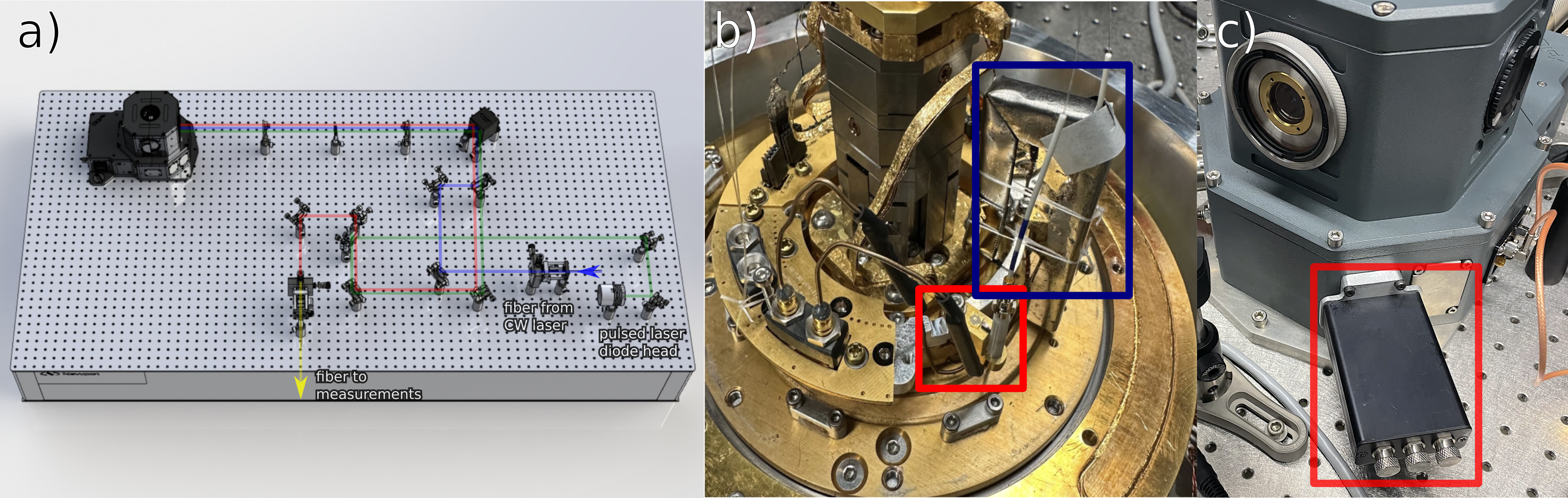}
    \caption{The ICECAP system and confocal microscopy. a) The 4f confocal microscope with 2D scanning capabilities built in front of the cryostation. Excitation can originate from a 785 nm CW laser (blue) or a 1080 nm picosecond pulsed laser (green). The collection path (red) is fiber-coupled and then can either connect to a spectrometer or the SNSPDs in the cryostat. b) Internal photo of sample space and the crux of the ICECAP system. The 1.56 K base temperature platform is the raised center podium. The SNSPDs are boxed in red and mounted directly onto the base platform to maximize thermal contact. They are situated in the shadow of the nanopositioner stack to reduce noise from laser light scattering enough to allow time-correlated single-photon counting measurements, but additional radiation shielding can decrease noise further (boxed in dark blue), as long as it is well-thermalized to the 1.56K stage. The pigtailed fiber from the feedthrough is attached from above and can be carefully exchanged with pigtailed fiber from the sample stage to avoid table-top collection routing. c) An external photo of the vacuum shroud with modifications for an in-vacuum low working distance microscope objective (top-left) and a custom-built fiber feedthrough (boxed in red).}
    \label{fig:Fig1}
\end{figure}

Traditionally, a 4f confocal microscope for single-photon cryogenic measurements of nanophotonic systems, like the one illustrated in Figure~\ref{fig:Fig1}a, consists of a few core parts: a pump laser, steering optics, the 4f optics\footnote{4f optics consist of a mirror, two lenses each with focal length f, and a microscope objective. The mirror is placed as the starting point. The first lens is placed 1f away from the mirror and the second lens is placed 2f from the first lens. Finally the microscope objective is placed 1f from the second lens giving a total length of 4f.}, sample cryostat, and single-photon detectors. Current commercially available single-photon detectors, such as single-photon avalanche diodes, that are housed at room temperature typically have collection efficiencies at about $50 \%$ or lower and dark counts of $< 100$ Hz. These figures of merit can be improved upon by using SNSPDs. However, SNSPDs have typically worked at $T < 2$ K, which come with monetary costs of buying and maintaining an extra cryostat in addition to the time cost of maintaining and the physical lab space required to house the extra equipment. For optically accessible non-dilution cryostats, the typical operating temperatures have been $\geq 3$ K which has precluded incorporating SNSPDs directly into the optical cryostat. Optical dilution refrigerators exist, but they generally require microscope objectives with much longer working distances and lower numerical aperture, making single-photon collection more difficult.

The cryostat used in this work is a Montana Instruments Cryostation xp100 model with a 2-inch diameter sample space and a base temperature of 1.56 K. The microscope objective is internal to the vacuum shroud which allows for the use of an objective with numerical aperture of $0.85$ to maximize collection efficiency. The SNSPDs (Figure~\ref{fig:Fig1}b), fiber feedthrough (Figure~\ref{fig:Fig1}c), and SNSPD control electronics used are from Quantum Opus and are maximally efficient at 1310 nm with an operating temperature of 2.5 K or below. Further, the SNSPD used in this work has an instrument response function with a full width at half maximum of 170 ps, well below previously reported NV center lifetimes, so signal deconvolution is not necessary.

The sample in the cryostat is optically accessible without the use of fiber optics, which allows for relatively easy use of confocal microscopy techniques. Its low temperature capabilities also make prototyping on-chip integrated SNSPDs much easier and will make possible in-situ comparisons between on-chip experiments and commercially available SNSPDs. The sensitivity and recovery time of SNSPD devices built into the cryostat allow for fast experiments, like the lifetime measurements reported in this paper that comprised of millions of photon detections and were taken in a few hours. Further, this technique can also be expanded toward fully inside-the-cryostat detection by coupling light from a photonic device to a tapered fiber \cite{krumrein2024precise, PhysRevApplied.8.024026} and subsequently directly coupling the fiber to the SNSPD inside the chamber. 

\section{Optical measurements}
\begin{figure}[htpb]

    \centering
    \includegraphics[width =0.6\textwidth]{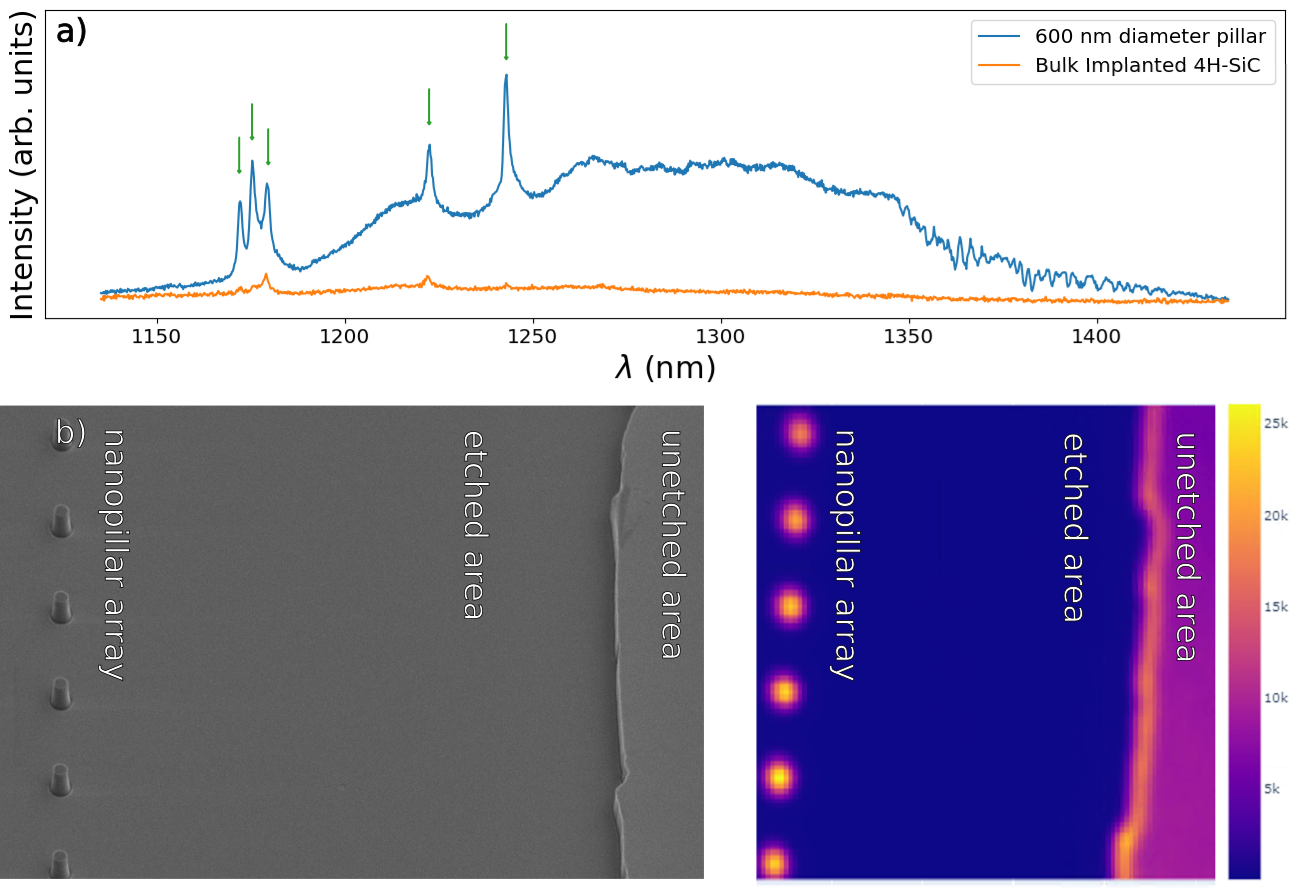}
    \caption{Measurements at cryogenic temperatures ($<$ 2 K) with a 785 nm laser at 5 mW power. a) PL spectrum of a 4H-SiC chip with integrated color centers. NV center ZPLs (1170 - 1245 nm) are marked with green arrows. b) Left: SEM image of a section on the chip with both fabricated nanopillar structures and an unetched area. Right: 2D PL scan of same section. Color bar signifies photon detection counts, yellow (blue) corresponds to more (less) counts.}
\label{fig:Fig3}

\end{figure}

The NV centers in the fabricated nanopillars are characterized by performing photoluminescence (PL) measurements in ICECAP. First, the presence of the four zero-phonon lines (ZPLs) corresponding to the four inequivalent lattice sites of the NV center in 4H-SiC is confirmed. Figure~\ref{fig:Fig3}a has the results of one such measurement performed at 1.56 K with a pump laser of 785 nm; the ZPLs agree with the literature values \cite{koehl2011room, falk2013polytype, Zargalah_ZPL_NV, PhysRevB.92.064104}. For all of the measurements reported in this work, a longpass filter (1150 nm cut-off) is used to remove the signal from the intrinsic divacancy color centers. 

A test of the SNSPDs is undertaken by focusing on the part of the chip that has both fabricated nanopillars (in this case 600 nm diameter pillars) and an unetched area; a 2D PL scan is performed across it. This area is selected to minimize possible variations in defect density across the whole sample and any variations in height that may affect the depth of focus for the excitation and collection optics allowing for a direct comparison between bulk and nanopillar measurements. Further, as the beam spot size is 1.2 $\mu$m and the density of NV centers across this area of interest is constant, we can consider the number of NV centers excited in bulk and in the 600 nm diameter nanopillars to be of the same order. Figure~\ref{fig:Fig3}b shows that not only are the single-photon detectors detecting signal from the PL scan, but also that there appears to be enhanced collection from the pillars compared to the unetched bulk, as expected from the simulations reported in the previous section. The completely etched area in between these two regions of interest shows no light collection because the etching depth is greater than the color center implantation depth, so any NV centers that were present have been etched away. 

\begin{figure}
    \centering
    \includegraphics[width=\linewidth]{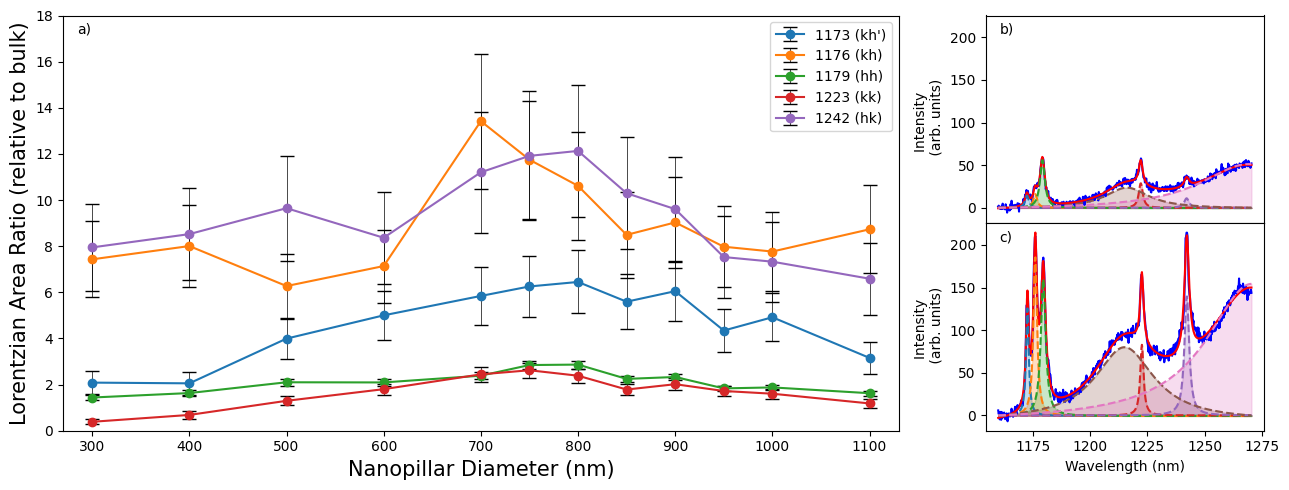}
    \caption{a) The collection enhancement ratio for a range of nanopillar diameters. Each PL spectrum is fit as a sum of seven Lorentzians, one for each of the five peaks, and two for the two distinct phonon sidebands visible in the spectrum. Colors correspond to Lorentzian fits from spectra in b) and c). The collection enhancement is then reported as the ratio of the area of the Lorentzian extracted from the nanopillar PL to the area of the Lorentzian extracted from the bulk PL measurement. There is a clear difference between the enhancement from the axial and basal orientations with the basal orientations outperforming the axial ones by a factor of ~6. b) The PL spectrum collected from bulk; note the axial peaks (red, green) are significantly higher than the basal (blue, orange, purple). c) The PL spectrum collected from an 800 nm diameter nanopillar; note the overall intensity enhancement and, in particular, that the basal orange and purple peaks are now outperforming the axial peaks.}
    \label{fig:Fig4}
\end{figure}

To check that the enhanced collection mentioned above includes photons emitted from NV centers, we compare measured NV center ZPL photon counts extracted from the PL spectra from the nanopillars and from the bulk unetched area shown in Figure~\ref{fig:Fig3}a.  The fitting method is shown in the inset of Fig.~\ref{fig:Fig4}. The areas of the Lorentzians of each peak are then compared to those measured from the bulk sample. The increase in areas for each ZPL can be seen in Fig.~\ref{fig:Fig4}. There is statistically significant collection enhancement for all 5 peaks measured.  The axial ZPL counts (1179 nm and 1222 nm) are enhanced about two-fold, while the basal ZPLs (1176 nm and 1242 nm) achieve 8- to 14-fold collection efficiency enhancement. The 1173 nm peak (along with the 1176 nm peak) was initially misclassified as being due to a tungsten impurity \cite{Zargalah_ZPL_NV}, but further studies have ruled out the tungsten lines \cite{PhysRevB.98.195202} and confirmed via EPR that the 1176 nm peak is due to the $kh$ orientation of the NV \cite{EPRtungsten}.
The origin of the 1173 nm peak is still not adequately explained, but polarization studies strongly suggest it arises from the $kh$-oriented NV \cite{PhysRevB.109.235203}, possibly as a result of residual strain \cite{ivanov2023near}.

These enhancement measurements demonstrate the trend seen in Fig~\ref{fig:Fig2}a-b in which simulated horizontally oriented dipoles emit more light out-of-plane than vertically oriented dipoles in bulk, but the reduction of in light emitted into bulk vs light emitted out-of-plane when integrated into nanopillars is much greater in the vertical orientation. As the axial site dipoles are horizontally oriented, while the basal site dipoles are oriented between $19^\circ$ and $90^\circ$, it is expected that basal NV centers will experience higher collection efficiency enhancement than axial centers, when placed into a pillar.

\begin{figure}
    \centering
    \includegraphics[width=\linewidth]{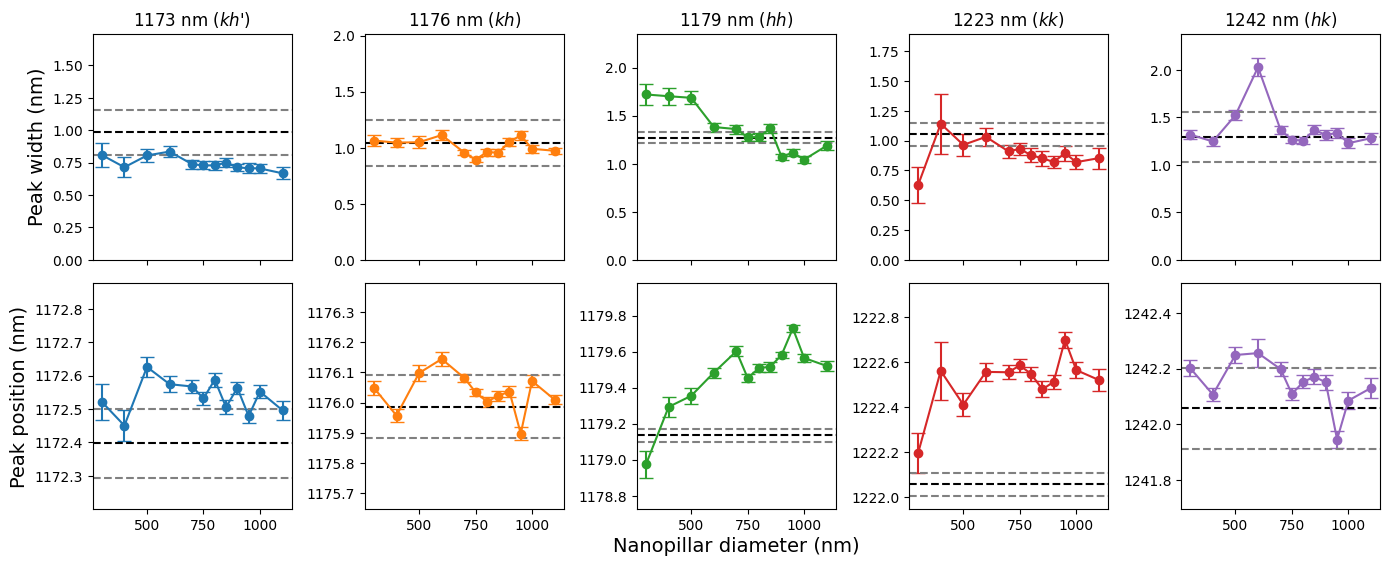}
    \caption{Peak positions and widths of nanopillars with a range of diameters relative to bulk measurements. Values reported are extracted from the same fits described in Fig~\ref{fig:Fig4}. In these plots, the thick black dashed line is the bulk measurement and the gray dashed lines are its error in fitting parameters. The top plots measure ZPL width. This is essentially a measure of the quality of the color centers as photon indistinguishability metrics are dependent on peak widths. Broadening of ZPLs occurs during fabrication due to damage in the crystal structure and can effectively rule out large-scale photonic integration. Very little spectral broadening is demonstrated in these plots. The bottom plots measure peak position away from the bulk measurement value. This is another method of exploring crystal damage as surrounding inhomogeneous strains couple to the emission wavelength. If more strain was being induced as nanopillar diameter decreased, peak position would be expected to change rapidly with diameter decrease. As this does not occur, we conclude that any strain induced is negligible.}
    \label{fig:Fig5}
\end{figure}

Another important conclusion drawn from the PL spectra comparison study is the extent to which fabrication-induced strain and surface effects affect the spectral properties of the ZPLs, which, while minor in both cases, we find is more pronounced for axial defects. Figure~\ref{fig:Fig5} reports the ratio of the widths and center wavelengths of each of the Lorentzians calculated from the nanopillar PL spectra to those calculated from the bulk spectrum. It is evident that the axial defects stray further from the bulk measurements than the basal defects in both peak wavelength and peak width while the basal defect measurements stay generally within one standard deviation of the bulk measurement. However, the highest fabrication-to-bulk ratio is only a factor of 2 at 1242 nm, which amounts to an increase of less than one meV.

Additionally, by observing the trends of spectral characteristics for which there is a significant difference between axial and basal orientations, the basal orientation of the 1173 nm peak (blue) is further supported. Not only does it follow the collection enhancement trend of the of the basal sites, which could be partially explained by the polarization reported in Ref. \cite{PhysRevB.109.235203}, but also the strain effects demonstrated in Fig~\ref{fig:Fig5} by the peak position. While the peak position of the 1173 nm line deviates beyond the fitting error of that of the bulk measurement, which demonstrates greater strain susceptibility, it still maintains the same trend as the confirmed $kh$ and $hk$ peaks. This minor discrepancy between the confirm may also point toward the 1173 nm peak being a product of substrate strain.

\begin{figure}
    \centering
    \includegraphics[width = \textwidth]{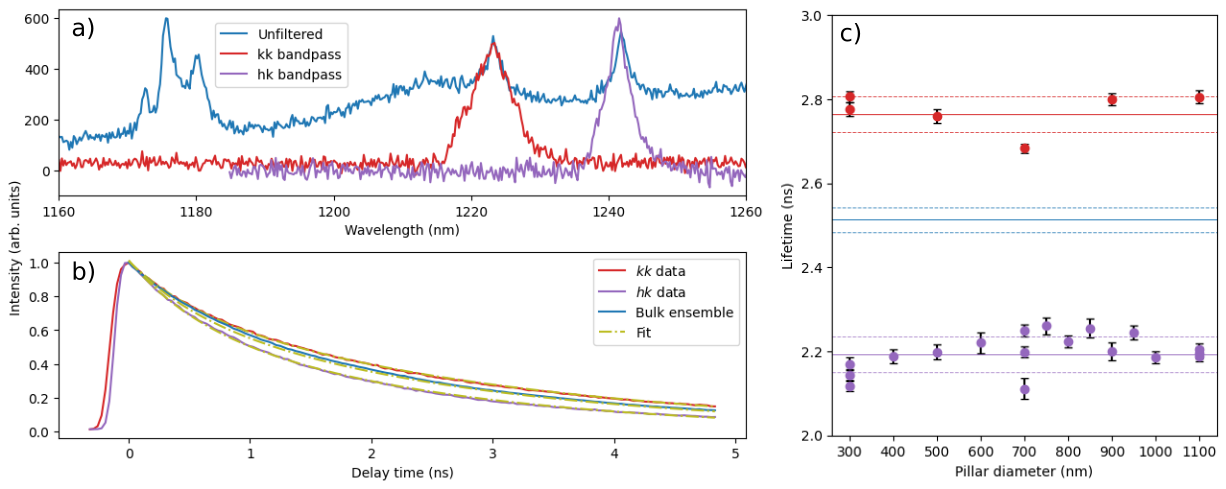}
    \caption{a) Photoluminescence spectrum of NV centers source (blue) and the same spectrum filtered by bandpass filters selecting $kk$ (red) and $hk$ (purple) ZPL wavelengths.
    b) The ensemble lifetime measurements from bulk (blue) and 600 nm diameter nanopillar and the Lorentzian fits (olive) for the $kk$ (red) and $hk$ (purple) orientations taken using the bandpass filters characterized in a). The measured lifetimes are $\tau_{kk} = (2.8 \pm 0.1)$ ns and $\tau_{hk} = (2.2 \pm 0.1)$ ns. 
    The acquisition time is 5 hours. The small non-linearity in the very short timescale when the lifetime is plotted on a log-lin plot is due to  the secondary short-lived decay from the backscatter of the laser pulse off of reflective elements in the sample chamber. The log-lin plots then become linear, indicating a single, dominating decay constant which is the fluorescence lifetime of the selected NV center. c) The lifetime measurements of nanopillars with a range of diameters taken using the $kk$ and $hk$ bandpass filters in a). The bulk measurements are denoted by solid lines with the dotted lines representing the corresponding fitting error. The significant difference between the $kk$ and $hk$ fluorescence lifetimes is present across nanopillar diameters. The lifetimes do not change with diameter, making a further case that these defects are spectrally robust to the fabrication methods used.}
    \label{fig:Fig6}
\end{figure}

Additionally, we measure the optical lifetime of all four NV center orientations collectively in the 600 nm diameter nanopillars. Due to the enhanced collection from the nanopillars samples, we are able to measure the ensemble lifetimes of the $kk$ and $hk$ orientations using bandpass filters formed from rotating both a long-pass filter (Thorlabs FELH1250) and a short-pass filter (Edmund Optics 89-675 1250 nm SP) under white light excitation until only the regions of interest are passed Fig.~\ref{fig:Fig6}a is a plot of the unfiltered PL spectrum from the 600 nm diameter pillars and plotted on top of it is the spectrum filtered using the $kk$ and $hk$ bandpass filters. Plots of the lifetime measurements are shown in Figure~\ref{fig:Fig6}. Each decay is fitted to a biexponential function $A_{1} \exp(-t/\tau_{1}) + A_2 \exp(-t/\tau_{2})$, where $A_{1}$, $A_{2}$, and $\tau_{1}$ are fitting parameters and $\tau_{2} \approx 0.54$ ns is fixed as the decay rate of the pump laser pulse reflecting off surfaces inside the sample chamber, which was measured separately. This is necessary because these back-reflections of pump light are simultaneous with color center emission. While this back-reflection signal is significantly attenuated by a 1150 nm longpass filter prior to detection, it is still present in the data due to its vastly greater initial brightness relative to the NV center emission. 

Measuring the 600 nm diameter pillars, we find a collective lifetime $\tau = (2.5 \pm 0.1)$ ns, which is in proximity to the previously reported value $\tau = (2.7 \pm 0.1)$ ns of ensemble NV center lifetimes measured at 20 K \cite{JFWangCoherentcontrol}. The individual ZPL lifetimes are $2.8 \pm 0.1$ ns and $2.2 \pm 0.1$ ns for the $kk$ and $hk$ ZPLs respectively. This disagrees slightly with the values reported for single NV centers at room temperature which range between 2.3 ns and 2.5 ns for three axial- and three basal-labeled NV centers; the specific orientation of the sites ($hk$ vs $kh$ or $kk$ vs $hh$) is not explicitly mentioned in the cited work \cite{JFWangExpOp}. To further explore the axial and basal discrepancy we observed in the measurements, we measured the individual ZPL lifetimes from a range of nanopillar diameters, as shown in Fig~\ref{fig:Fig6}d and e, and found the difference to bear out in both high (smaller diameter) and low (larger diameter) strain environments.

While non-radiative energy transitions in 4H-SiC have not been studied, some inference can be drawn by comparing the lifetimes measured from the 600 nm diameter nanopillars and from the pristine area. We find the bulk collective lifetime as measured from the 600 nm nanopillars to be ($2.85 \pm 0.04$) ns, and individual ZPL lifetimes of ($2.76 \pm 0.1$) ns and ($2.2 \pm 0.1$) ns for the $kk$ and $hk$ ZPLs respectively. The individual ZPL lifetimes from the bulk agreed with the measurements from the nanopillars. While the collective lifetime differed slightly between bulk and nanopillar measurements, it is similarly close to the previously reported higher temperature bulk measurements referenced above. That the lifetimes are similar lends evidence to the claim that the collection enhancements of the ZPLs is not due to factors like higher quantum efficiency due to stronger radiative recombination effects in the nanopillars and is more likely due to waveguiding effect of the nanopillars.

\section{Discussion}

In this work we demonstrate, to our knowledge, the first integration of 4H-SiC NV$^-$ centers into nanopillars and report the spectral properties of  NV center ensembles at temperatures below 2 K. These nanopillars display enhanced collection, up to 14-fold, relative to bulk measurements which will enable SiC NV center applications as efficient single photon sources and spin-photon entangling interfaces. Moreover, this enhancement is differentiating for axial versus basal color centers and, in addition to being a practical tool for higher sensitivity measurements, it can serve as an assessment tool for the identification of the structure of color centers and agreements with theoretical calculations of their dipole orientations. The measured lifetimes and spectral properties agree well with literature results for both ensemble and single NV centers at higher temperatures in bulk samples, and no significant spectral broadening from nanofabrication-induced strain or surface proximity is observed. Notably, careful analysis of PL spectra and lifetime measurements has led to some interesting questions about how the defect site symmetry affects spectral properties. The $hk$ site defect in particular, with its shorter fluorescence lifetime, robustness to fabrication stress, proximity to O-band wavelengths and predicted long coherence times\cite{PhysRevMaterials.5.074602} is a highly promising candidate for networking applications. The success of these measurements paves the way for more complex nanophotonic structures like on-chip beamsplitters or integration with on-chip single-photon detectors \cite{majety2023triangular, majety2023triangular2}. 

In this work we also developed new instrumentation, named ICECAP, that incorporates SNSPDs directly into the cryogenic sample chamber for spectroscopic characterization of NIR quantum emitters. The versatility of this system should be of particular interest to industry partners. Turn-key ICECAP instruments could conceivably be produced for less than the cost of a separate optical cryostat and SNSPD cryostat combination. In terms of building the future quantum workforce, movement toward more monolithic systems would lead to technicians and engineers needing training on fewer instruments. Integrated instruments like ICECAP will also lead to a  decrease in energy, helium, and total physical footprint costs of future QIST centers. Furthermore, the combination of the ICECAP system, refined fabrication techniques, and the favorable properties of NV centers in 4H-SiC as spin-1 color centers that emit single photons in the near infrared makes for a rapidly emerging quantum information technology and science workhorse.

\begin{acknowledgement} We acknowledge support from NSF CAREER (Award 2047564). Work at the Molecular Foundry was supported by the Office of Science, Office of Basic Energy Sciences, of the U.S. Department of Energy under Contract No. DE-AC02-05CH11231. This work was partially supported by the UC Davis Physics REU program under NSF grant PHY2150515.

The authors acknowledge Dr. Tevye Kuykendall from the Inorganic Nanostructures Facility at the Molecular Foundry for his valuable support with the annealing furnaces. We thank Amy Conover and Dr. Aaron Miller from Quantum Opus and Ian Durnford and Taylor Bohach from Montana Instruments for their support when incorporating the SNSPDs and cryostat into a single system. Part of this study was carried out at the UC Davis Center for Nano and Micro Manufacturing (CNM2). 
\end{acknowledgement}

\bibliography{main}

\end{document}